\documentclass[twocolumn,aps,prl] {revtex4} %% 2-spaltig
\usepackage{graphicx}
\begin{document}
\pagestyle{empty} 
\title{Interfacial separation between elastic solids with randomly rough surfaces: comparison of experiment with theory}
\author{B. Lorenz and B.N.J. Persson}
%\affiliation{Institut f\"ur Festk\"operForschung, Forschungszentrum J\"ulich, D-52425 J\"ulich, Germany}
\affiliation{IFF, FZ-J\"ulich, D-52425 J\"ulich, Germany}

\begin{abstract}
We study the average separation between an elastic solid and a hard solid 
with a nominal flat but 
randomly rough surface, as a function of the squeezing pressure. We present experimental
results for a silicon rubber (PDMS) block with a flat surface squeezed against an asphalt
road surface. The theory shows that an effective 
repulse pressure act between the surfaces of the form
$p\sim {\rm exp}(-u/u_0)$, where $u$ is the average separation between the surfaces and
$u_0$ a constant of order the root-mean-square roughness, 
in good agreement with the experimental results.
\end{abstract}
\maketitle

%%%%%%%%%%%%%% main text %%%%%%%%%%%%%%%%

\vskip 0.3cm
{\bf 1 Introduction}

Contact mechanics between solid surfaces is the basis for understanding many
tribology processes\cite{MB,Bowden,Johnson,BookP,Isra,Capillary.adhesion,Sealing} 
such as friction, adhesion, wear and sealing. The two most important
properties in contact mechanics are the area of real contact and the interfacial separation between the
solid surfaces. 
For non-adhesive contact and small squeezing pressure, 
the (projected) contact area depends linearly on the 
squeezing pressure\cite{P1,Greenw,Muser,Carbone}.

When two elastic solids with rough surfaces are squeezed together, the solids will
in general not make contact everywhere in the apparent contact area,
but only at a distribution of asperity contact spots. The separation
$u({\bf x})$ between the surfaces will vary in a nearly random way with the lateral
coordinates ${\bf x}=(x,y)$ in the apparent contact area. 
When the applied squeezing pressure increases, the average 
surface separation $u=\langle u({\bf x})\rangle$ will decrease, but in most situations it is not
possible to squeeze the solids into perfect contact corresponding to $u=0$. 
We have recently developed a theory which predicts 
that, for randomly rough surfaces at low squeezing pressures, $p \sim {\rm exp} (-u/u_0)$,
where the reference length $u_0$ depends on the nature of 
the surface roughness but is independent of $p$\cite{MB,P4}.
Here we will present experimental results to test the theory predictions\cite{Kroger}.
We study the squeezing of a rubber block against an asphalt road surface. This topic is also
important in the context of the air-pumping contribution to tire noise\cite{airpump}. Thus the
compression and outward flow of the air between a tread block and the road surface cavities 
during driving contribute to tire noise, similarly to 
how sound is generated during applause. A similar effect 
(but now involving decompression and inward flow of air)
occur when a tread block leave the tire-road contact area.

\begin{figure}
\includegraphics[width=0.45\textwidth,angle=0]{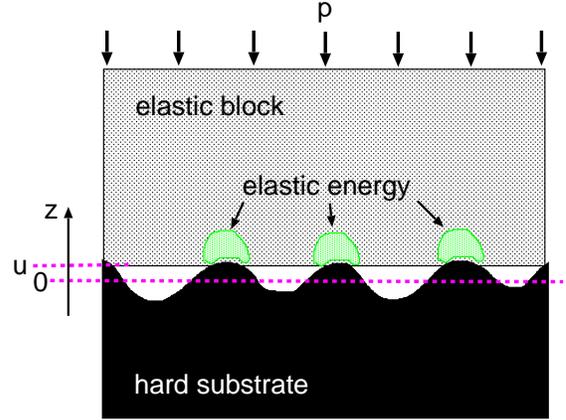}
\caption{\label{block}
An elastic block squeezed against a rigid rough substrate. The separation
between the average plane of the substrate and the average plane of the lower
surface of the block is denoted by $u$. Elastic energy is stored in the block in the vicinity
of the asperity contact regions.
}
\end{figure}

\vskip 0.3cm
{\bf 2 Theory}

We consider the frictionless contact between an elastic solid (elastic modulus
$E$ and Poisson ratio $\nu$) with a flat surface and a
rigid, randomly rough surface with the surface height profile $z=h({\bf x})$. 
The separation between the average surface plane of the block and the
average surface plane of the substrate (see Fig. \ref{block}) is denoted by $u$ with $u \ge 0$.
When the applied squeezing force $p$
increases, the separation between the surfaces at the interface will decrease, and
we can consider $p=p(u)$ as a function of $u$. 
The elastic energy $U_{\rm el}(u)$ stored in the substrate asperity--elastic block contact regions
must be equal to the work done by the external pressure $p$ in displacing the lower surface of the
block towards the
substrate. Thus,
$$p(u) = - {1\over A_0} {d U_{\rm el} \over du},\eqno(1)$$ 
where $A_0$ is the nominal contact area. Eq. (1) is exact\cite{P4,YP}.
Theory shows that for low squeezing pressure, the area of real contact $A$ varies linearly with
the squeezing force $pA_0$, and that the interfacial stress distribution, and the 
size-distribution of contact spots, are independent of the squeezing pressure\cite{P3}. 
That is, with increasing $p$
existing contact areas grow and new contact areas form in such a way that in the thermodynamic limit
(infinite-sized system) the quantities referred to above remain unchanged. It follows immediately
that for small load {\it the elastic energy stored in the asperity contact region will increase
linearly with the load}, i.e., $U_{\rm el}(u) = u_0 A_0 p(u)$, where $u_0$ is a characteristic length
which depends on the surface roughness
(see below) but is independent of the squeezing pressure $p$. Thus, for small pressures (1) takes the form
$$p(u) = - u_0 {d p \over du}$$
or\cite{other}
$$p(u) \sim e^{- u/u_0}.\eqno(2)$$ 

To quantitatively derive the relation $p(u)$ we need an analytical expression for the asperity
induced elastic energy. Within the contact mechanics approach of Persson 
we have\cite{P3,P2,CARL}
$$U_{\rm el} \approx A_0 E^* {\pi \over 2} \int_{q_0}^{q_1} dq \ q^2 P(q,p)C(q),\eqno(3)$$
where $E^*=E/(1-\nu^2)$ and
where $P(q,p)= A(\zeta) /A_0$ is the relative contact area when the interface is studied at
the magnification $\zeta = q/q_0$, which depends on the applied pressure $p$. 
The surface roughness power spectrum\cite{P3}
$$C(q)= {1\over (2\pi )^2} \int d^2x \langle h({\bf x}h({\bf 0})\rangle
e^{-i{\bf q}\cdot {\bf x}},$$
where $\langle..\rangle$ stands for ensemble average.  
Note that for complete contact $P=1$ and in this limit (3) is exact.
For self affine fractal surfaces the prediction of the contact mechanics theory of Persson 
has been compared to numerical simulations\cite{CARL,Hyun}. 
The numerical studies indicate that as the fractal dimension
of the surface approaches 2 the Persson theory may become exact, while a small difference between theory
and simulations are observed for larger fractal dimension\cite{short}. 
Below we will compare the theory predictions
to experimental data for an asphalt road surface which is fractal-like with the fractal dimension $D_{\rm f}
\approx 2$. We find nearly perfect agreement between theory and experiment (see below), supporting the picture
gained before based on numerical simulations. 

Substituting (3) in (1) gives for small squeezing pressures\cite{P4}:
$$p={\beta E^*} e^{-u/u_0}\eqno(4)$$
For self affine fractal surfaces,
the length $u_0$ and the parameter $\beta$ depend on the Hurst exponent
$H$ and on $q_0$ and $q_1$. 
Most surfaces which are self affine fractal have the Hurst exponent $H > 0.5$
(or the fractal dimension $D_{\rm f} < 2.5$). For such surfaces $u_0$ and $\beta$ 
are nearly independent of the highest 
surface roughness wavevector, $q_1$, included in the analysis. For the substrate surface studied 
below we obtain from the measured surface roughness power spectrum (see Fig. \ref{logq.logC.old.asphalt})
$u_0 = 0.30 \ {\rm mm}$ and $\beta = 0.59$. Note that $u_0$ is of order the root-mean-square
roughness amplitude ($h_{\rm rms} \approx 0.29 \ {\rm mm}$ in the present case, see below). 

\begin{figure}
\includegraphics[width=0.47\textwidth,angle=0]{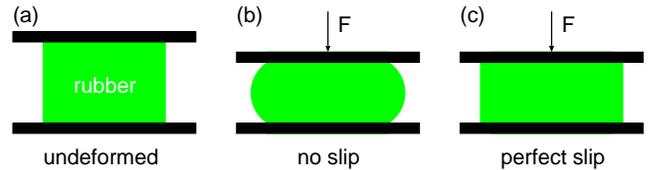}
\caption{\label{threeblock}
A rubber block between two flat and rigid solid plates. 
(a) Undeformed state. (b) Squeezed block assuming no slip (i.e., high enough static friction)
at the rubber-plate
interfaces. (c) Squeezed block assuming perfect slip (i.e., no friction) at the rubber-plate
interfaces.}
\end{figure}

\begin{figure}
\includegraphics[width=0.55\textwidth,angle=0]{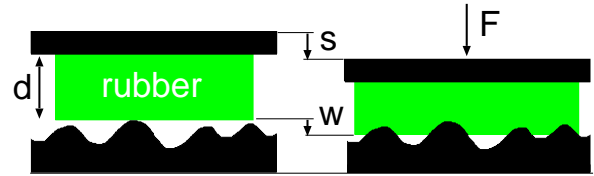}
\caption{\label{compressed}
A rubber block in contact with a rigid, randomly rough substrate. Left: no applied load.
Right: The rubber block is squeezed against the substrate with the force $F$. The upper and the 
(average position of) the lower surface of the rubber block moves downwards by the distances $s$
and $w$, respectively. We assume perfect interfacial slip (no friction).} 
\end{figure}

Consider a rubber block (elastic modulus $E$) with a flat surface (area $A_0$) and the thickness $d$. 
We will study both dry and lubricated interfaces
(see Fig. \ref{threeblock}) resulting in no slip and perfect slip at the two rubber-confining wall
interfaces.
If the block is squeezed against a rigid, randomly rough counter surface,
the upper surface of the rubber block will move downwards by the distance $s$ 
(see Fig. \ref{compressed}), which is the sum of a uniform
compression of the rubber block, $d \sigma/E$, and a movement (or penetration)
$w$ of the average position of the 
lower surface of the rubber block into the valleys or cavities of the countersurface:
$$s=w+d \sigma/E\eqno(5)$$
If $u$ denote the average separation between the block and the substrate (so that $u = 0$
correspond to perfect contact) then
$$w=h_{\rm max}-u\eqno(6)$$
where we have assumed that the initial position of the lower surface of the block correspond to
the separation where the block just makes contact with the highest substrate asperity (as in Fig.
\ref{compressed}, left), which is located
a distance $h_{\rm max}$ above the average substrate surface plane. 
Using (4) we get
$${\rm log} (\sigma /E) = {\rm log}( 4\beta /3) -u /u_0\eqno(7)$$ 
where $\sigma = F/A_0$ the
squeezing pressure. Here we have used that $E^*/E=1/(1-\nu^2) \approx 4/3$ since for rubber
$\nu \approx 1/2$. 
Combining (5) and (6) gives
$$u = h_{\rm max}-s+d \sigma/E$$
Substituting this in (7) gives
$${\rm log} \left ({\sigma \over E}\right ) = {\rm log}\left ( {4 \beta\over 3}\right ) -{1\over u_0} 
\left (h_{\rm max}-s+d {\sigma \over E} \right )$$ 
or 
$${\rm log} \left ({\sigma \over E}\right ) = B+{1\over u_0}\left (s-d {\sigma \over E} \right )\eqno(8)$$
where $B={\rm log}(4 \beta/ 3) -h_{\rm max}/u_0$. 

For no-slip boundary condition, Eq. (5) is replaced by
$$s=w+d \sigma/E'$$
where the effective modulus $E'>E$. Thus, in this case (8) takes the form
$${\rm log} \left ({\sigma \over E'}\right ) = B'+{1\over u_0}\left (s-d {\sigma \over E'} \right )\eqno(9)$$
where
$B'={\rm log}(4 \beta E/ 3E') -h_{\rm max}/u_0$.
\vskip 0.3cm
{\bf 3 Experimental}

To test the theory presented above, we have performed the experiment indicated in Fig. \ref{compressed}. 
A rubber block with a flat surface was squeezed against an asphalt road surface. 
The displacement \textit{s} of the  upper surface of the rubber block 
was changed in steps of $0.05 \ {\rm mm}$, and the force \textit{F} was measured. For the experiment 
we used a test stand produced by SAUTER GmbH (Albstadt, Germany), normally used to 
measure spring constants. Using this test stand, we are able to measure  
forces up to $500 \ {\rm N}$, and displacement with the resolution $0.01 \ {\rm mm}$.

The rubber block was made from a silicone elastomer (PDMS). 
The PDMS samples were prepared using a two-component kit (Sylgard 184) 
purchased from Dow Corning (Midland, MI). This kit consists of a base 
(vinyl-terminated polydimethylsiloxane) and a curing agent 
(methylhydrosiloxane-dimethylsiloxane copolymer) with a suitable catalyst. 
From these two components we prepared a mixture of 10:1 (base/cross linker) in weight. 
The mixture was degassed to remove the trapped air induced by stirring from the mixing 
process and then poured into cylindric casts (diameter $D = 3 \ {\rm cm}$ and 
height $d = 1 \ {\rm  cm}$). 
The bottom of these casts were made from glass to obtain smooth surfaces (negligible roughness). 
The samples were cured in an oven at $80 \ ^\circ{\rm C}$  for over 12 hours.

The road surface used in this experiment was provided by Pirelli (Italian tire manufacturer). 
The topography was measured with contact-less optical methods using a chromatic sensor 
with two different optics produced by Fries Research \& Technology GmbH (Bergisch Gladbach, Germany). 
To identify the elastic modulus $E$, the PDMS sample was first squeezed against a smooth substrate 
in a compression test. We measured the force \textit{F} over the displacement \textit{s} 
for two different cases. First there was no lubrication used and the PDMS sample 
deformed laterally at the force-free area as shown in Fig. 2(b), because no slip 
occurred at the contact areas.
Second we lubricated the contact areas to obtain perfect slip at the 
interfaces (see Fig. 2(c)). We used polyfluoroalkylsiloxane (PFAS), a fluorinated silicone 
oil supplied by ABCR GmbH \& Co. KG (Karlsruhe, Germany). Because of its high viscosity  
($\eta = 1000 \ {\rm cSt}$), the fluid is an excellent lubricant also under extreme pressure 
applications and should therefore not easily be squeezed out of the contact area. 
Also it does not react (or interdiffuse) with the PDMS elastomer.

\begin{figure}
\includegraphics[width=0.47\textwidth,angle=0]{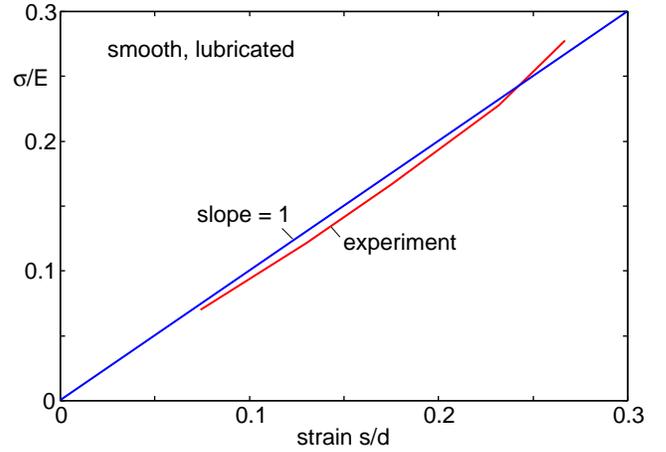}
\caption{\label{flat.lubricated.1u.over.d.2sigma.over}
The stress $\sigma$ (in units of the elastic modulus $E$) as a function of the strain $s/d$,
where $s$ is the displacement of the upper surface and $d$ the thickness of the block.
In the calculation we used $E=2.3 \ {\rm MPa}$.
For a PDMS rubber block confined between two smooth lubricated (wet) surfaces.}
\end{figure}

\begin{figure}
\includegraphics[width=0.47\textwidth,angle=0]{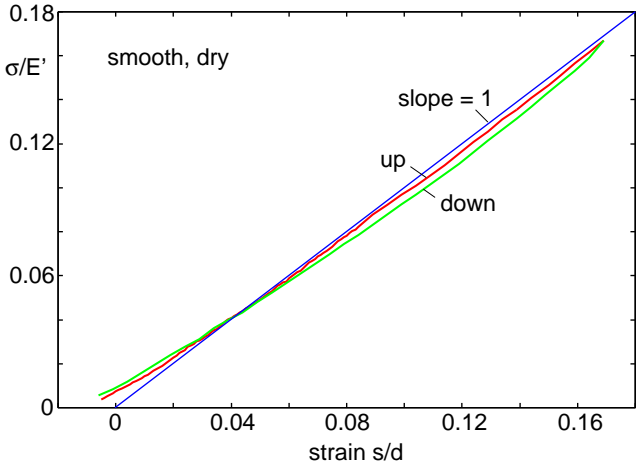}
\caption{\label{FLAT.1u.over.d.2sigma.over.E}
The stress $\sigma$ (in units of the elastic modulus $E'$) as a function of the strain $s/d$,
where $s$ is the displacement of the upper surface and $d$ the thickness of the block.
In the calculation we used the effective modulus $E'=4.2 \ {\rm MPa}$.
For a PDMS rubber block confined between smooth dry surfaces. The two experimental
curves corresponds to increasing and decreasing the strain.}
\end{figure}

\begin{figure}
\includegraphics[width=0.47\textwidth,angle=0]{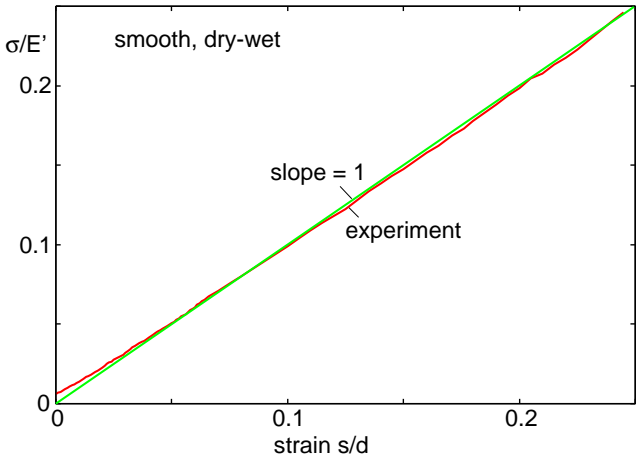}
\caption{\label{one.wet.one.dry.FLAT}
The stress $\sigma$ (in units of the elastic modulus $E'$) as a function of the strain $s/d$,
where $s$ is the displacement of the upper surface and $d$ the thickness of the block.
In the calculation we used the effective modulus $E'=2.9 \ {\rm MPa}$.
For a PDMS rubber block confined between one lubricated (wet) surface and one dry surface.}
\end{figure}

\vskip 0.3cm
{\bf 4 Results}

Consider first flat surfaces. In Fig.
\ref{flat.lubricated.1u.over.d.2sigma.over} we show the measured relation between the stress and the
strain for lubricated surfaces (so that the shear stress vanish on the boundaries). 
If the stress is normalized with $E=2.3 \ {\rm MPa}$ a nearly strait line with the slope $1$
will result so that the relation $\sigma = E s/d$ holds. The elastic modulus $E=2.3 \ {\rm MPa}$
is consistent with the elastic modulus reported in the literature for 
similar silicon rubbers\cite{Bon}. 

We have also performed experiments for dry surfaces. In this case no (or negligible) slip occur
at the interface with the confining walls, and visual inspection of the system showed that the
rubber bulge laterally at the force-free area (see Fig. \ref{threeblock}(b)). We
still expect a linear (or near linear) relation between stress and strain but the 
effective elastic modulus $E'$ is larger than for lubricated interfaces. 
Thus, the effective elastic modulus deduced 
from the experimental data (see Fig. \ref{FLAT.1u.over.d.2sigma.over.E}) 
$E' \approx 4.2 \ {\rm MPa}$ is about $80\%$ larger than for the
lubricated interface. 
To check the measuring system for hysteresis effects, some of the experiments were 
performed bidirectional. The results are shown in Fig. \ref{FLAT.1u.over.d.2sigma.over.E} 
where the strain was increased and after that 
slowly decreased again. 
Negligible hysteresis occur, as expected because of the low glass transition temperature of PDMS.

The increase in the effective elastic modulus in compression, from $2.3 \ {\rm MPa}$
to $4.2 \ {\rm MPa}$, when going from slip to no-slip
boundary condition, is consistent with the prediction of the Lindley equation\cite{Lin}, which
in the present case takes the form
$$E'\approx E\left (1+1.4 S^2\right)$$
For a cylinder the {\it shape factor} $S=R/2d$. In the present case $E= 2.3 \ {\rm MPa}$
and $S=0.75$ giving
$E'=4.1 \ {\rm MPa}$ which agree very well with the measured value ($4.2 \ {\rm MPa}$).

We have also studied the case where one surface is lubricated and the other dry. In this case
the the rubber will displace laterally in an assymetric way (as in Fig. \ref{twopress}(b)) and 
the measured effective elastic modulus $E'= 2.9 \ {\rm MPa}$ (see Fig. \ref{one.wet.one.dry.FLAT}), 
is slightly smaller than the
the average of the effective $E$-modulus obtained assuming
no-slip and complete slip on both surfaces: $(2.3+4.2)/2 \  {\rm MPa} \approx 3.3 \ {\rm MPa}$.

We will now present experimental results for a rubber block squeezed against an asphalt road surface.
The surface roughness power spectrum of the road surface is shown in Fig. \ref{logq.logC.old.asphalt}.
The surface has the root-mean-square roughness $h_{\rm rms} \approx 0.29 \ {\rm mm}$, 
and for the wave vector $q>q_0
\approx 2500 \ {\rm m}^{-1}$ it is (on a log-log scale) 
well approximated by a strait line with the slope corresponding to a self-affine fractal
surface with the fractal dimension $D_{\rm f}=2$. For $q<q_0$, $C(q)$ is approximately constant;
we refer to $q_0$ as the roll-off wavevector.

\begin{figure}
\includegraphics[width=0.47\textwidth,angle=0]{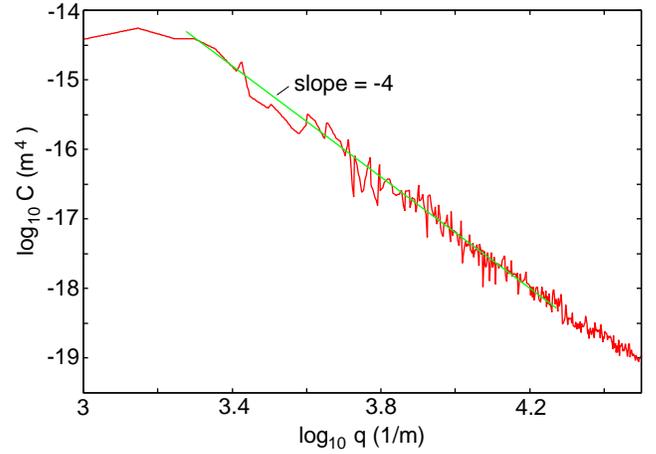}
\caption{\label{logq.logC.old.asphalt}
The surface roughness power spectrum $C$, as a function of the wavevector
$q$ (log-log scale), for an asphalt road surface. The strait green line has the slope $-4$,
corresponding to the Hurst exponent $H=1$ (fractal dimension $D_{\rm f}=2$).} 
\end{figure}

\begin{figure}
\includegraphics[width=0.47\textwidth,angle=0]{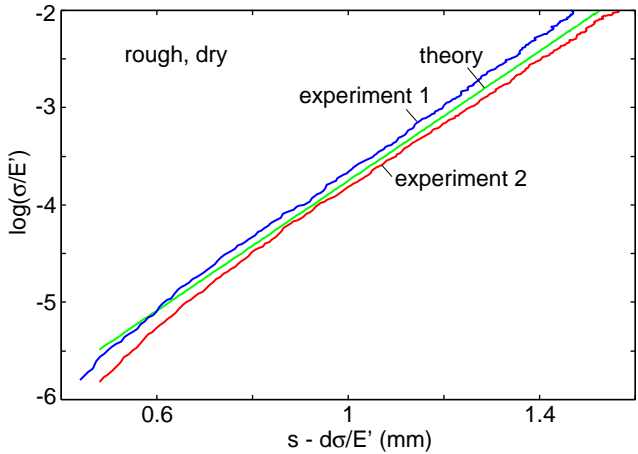}
\caption{\label{1u.minus.uflat.2ln.sigma.over.E}
The natural logarithm of the squeezing pressure (divided by the effective elastic modulus) as a 
function of $s-d\sigma/E'$, where $s$ is the displacement of the upper surface of the rubber block
relative to the substrate, and where $d$ is the thickness of the rubber block.
In the calculation we used the effective elastic modulus $E'=4.8 \ {\rm MPa}$
and $B'=-6.85$. For dry contact.}
\end{figure}

In Fig. \ref{1u.minus.uflat.2ln.sigma.over.E} we show 
the natural logarithm of the squeezing pressure (divided by the effective elastic modulus) as a 
function of $s-d\sigma/E'$, where $s$ is the displacement of the upper surface of the rubber block
relative to the substrate, and where $d$ is the thickness of the rubber block.
In the calculation we used the effective elastic modulus $E'=4.8 \ {\rm MPa}$ and
$B'=-6.85$. The  value of $B'$ has been calculated using (9) (using the measured $h_{\rm max}$)
so that the only fitting parameter is the effective elastic modulus $E'$, which however
agree rather well with the measurements for flat surfaces ($E'=4.2 \ {\rm MPa}$).

\begin{figure}
\includegraphics[width=0.47\textwidth,angle=0]{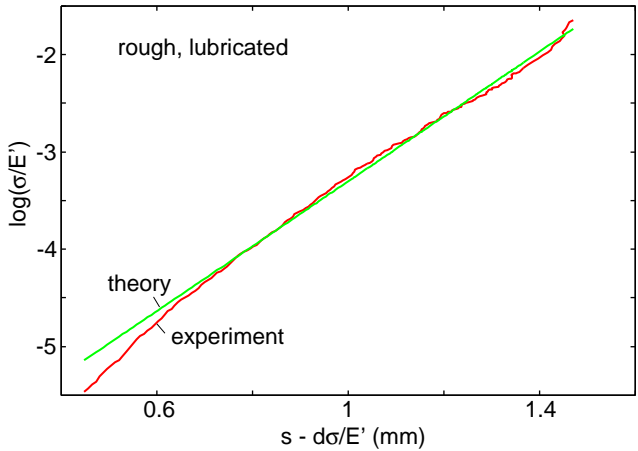}
\caption{\label{lubricated.u.minus.ublock.sigma.over.E}
The natural logarithm of the squeezing pressure (divided by the effective elastic modulus) as a 
function of $s-d\sigma/E'$, where $s$ is the displacement of the upper surface of the rubber block
relative to the substrate, and where $d$ is the thickness of the rubber block. For lubricated (wet) 
contact.
In the calculation we used the effective elastic modulus 
$E'=3.4 \ {\rm MPa}$ and $B'=-6.5$.}
\end{figure}

\begin{figure}
\includegraphics[width=0.40\textwidth,angle=0]{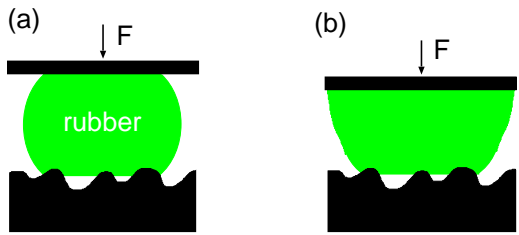}
\caption{\label{twopress}
A rubber block squeezed between a rigid solid plate and a rigid randomly rough substrate. 
(a) Dry surfaces and (b) lubricated surfaces.}
\end{figure}

In Fig. \ref{lubricated.u.minus.ublock.sigma.over.E} we show the same as in Fig. 
\ref{1u.minus.uflat.2ln.sigma.over.E} but now for lubricated surfaces.
In the calculation we used the effective elastic modulus 
$E'=3.4 \ {\rm MPa}$ and $B'=-6.50$. Note that this value for $B'$ is slightly smaller than for
dry contacts. The difference $\Delta B' = -6.50-(-6.85) = 0.35$ just reflect the difference in the
effective $E$-modulus since according to (9) $\Delta B' = {\rm log}[E'({\rm dry})/E'({\rm lubricated})]
={\rm log}(4.8/3.4) \approx 0.35$. The $E'$ value is larger than the $E$-modulus measured for flat
lubricated surfaces ($E=2.3 \ {\rm MPa}$), but this can be understood as follows.

Visual inspection of the contact between the 
rubber cylinder and the two confining walls shows that, as
expected from above,
the rubber block slip against the top (flat) steel surface, while no slip (or only very limited slip)
occur against the rough substrate surface, see Fig. \ref{twopress}(b). 
This is consistent with the fact that the observed
elastic modulus is larger than $E= 2.3 \ {\rm MPa}$, as obtained above when 
complete slip occur at both (lubricated) surfaces.
In fact, the observed effective $E$-modulus ($3.4  \ {\rm MPa}$) 
is rather close to the value $2.9 \ {\rm MPa}$ measured
for smooth surfaces when slip occur at one surface and no slip at the other surface.
The fact that no (or very small) slip occur at the interface between the rubber and the 
rough substrate surface may
be due to at least two facts:

1) The pressure in the asperity contact regions are much higher than the average pressure,
and the asperity contact regions much smaller than the nominal contact area, 
resulting in much faster squeeze-out
of the lubricant oil from the asperity contact regions, as compared to the case of flat surfaces, 
and consequently to higher friction in the contact 
regions. 

2) The substrate surface roughness on different length scales contribute to the friction during slip
because of the viscoelastic deformations of the rubber on different length scales. However,
since for silicon rubber viscoelastic dissipation only occur at very high frequencies, 
it is likely that this
effect is small in the present case.

The measured $E'$-values for rough surfaces ($4.8 \ {\rm MPa}$ and $3.4 \ {\rm MPa}$) 
are roughly $14 \%$ larger than for smooth
surfaces ($4.2 \ {\rm MPa}$ and $2.9 \ {\rm MPa}$), as obtained
assuming no-slip on the confining surfaces in one case, and slip on only one of
the confining surfaces in the other case. 
The origin of this (small) difference in effective elastic modulus is
not known to us.

Finally, we note that for $s-d\sigma /E' < 0.6 \ {\rm mm}$ the experimental curve in 
Fig. \ref{lubricated.u.minus.ublock.sigma.over.E} drops of faster with decreasing interfacial
separation than predicted by the theory.
(The same effect can also be seen in 
Fig. \ref{1u.minus.uflat.2ln.sigma.over.E} and has also been observed in 
molecular dynamics calculations\cite{YP}.)
This is a finite size effect:
The theory is for an infinite system which has (arbitrary many) arbitrary high asperities, and contact
between the two solids will occur for arbitrary large surface separation, and the relation 
$p\sim {\rm exp}(-u/u_0)$ holds for arbitrary large $u$. On the other hand a finite system has
asperities with height below some finite length $h_{\rm max}$, and for $u > h_{\rm max}$ no contact
occur between the solids and $p=0$.  

\vskip 0.3 cm
{\bf 5 Summary and conclusion}

We have presented a combined experimental-theoretical study of the contact 
between a rigid solid with a randomly rough surface and an elastic block with a flat surface. 
The interfacial separation as a function of the squeezing pressure has been 
derived theoretically and has been compared to the experimental results. 
We find nearly perfect agreement between theory and experimental data for an asphalt road surface. 
We conclude that for non-adhesive interaction and small applied 
pressure, $p \sim {\rm exp} (-u/u_0)$, where $p$ is the squeezing pressure and $u$ the average 
interfacial separation, and $u_0$ a constant of order the root-mean-square roughness of the combined surface
profile. In addition, the experimental results indicate that for surfaces with fractal-like roughness 
profiles the Persson contact mechanics theory may be exact for the fractal dimension $D_{\rm f}=2$.
We plan to extend the study above to surfaces with other fractal dimension to test the
theory in more general cases.
The presented results may be of great importance for, e.g., heat transfer, 
lubrication, sealing, optical interference, and tire noise related to air-pumping.

\vskip 0.3 cm
{\bf Acknowledgments}

The authors would like to thank Pirelli Pneumatici for support. We thank 
O.D. Gordan (IBN, FZ-J\"ulich) for help with preparing the PDMS rubber blocks.

\end{document}